\documentclass[prd,showpacs,preprintnumbers,amsmath,amssymb,superscriptaddress,floatfix,nofootinbib]{revtex4}
\usepackage{graphicx}
\usepackage{amsmath}
\usepackage{amsfonts}
\usepackage{amssymb}
\usepackage{color}
\usepackage{multirow}
\usepackage{footnote}

\begin{document}

\title{The scalar $f_0(500)$, $f_0(980)$, and  $a_0(980)$ resonances and vector mesons in the single Cabibbo-suppressed decays $\Lambda_c \to p K^+K^-$ and $p\pi^+\pi^-$ }

\author{Zhe Wang}
\affiliation{School of Physics and Microelectronics, Zhengzhou University, Zhengzhou, Henan 450001, China}

\author{Yan-Yan Wang}
\affiliation{School of Materials Science and Engineering, Zhengzhou
University of Aeronautics, Henan 450001, China}

\author{En Wang}
\email{wangen@zzu.edu.cn}
\affiliation{School of Physics and Microelectronics, Zhengzhou University, Zhengzhou, Henan 450001, China}

\author{De-Min Li}\email{lidm@zzu.edu.cn}
\affiliation{School of Physics and Microelectronics, Zhengzhou University, Zhengzhou, Henan 450001, China}

\author{Ju-Jun Xie}\email{xiejujun@impcas.ac.cn}
\affiliation{Institute of Modern Physics, Chinese Academy of
Sciences, Lanzhou 730000, China} \affiliation{School of Nuclear
Sciences and Technology, University of Chinese Academy of Sciences,
Beijing 101408, China} \affiliation{School of Physics and
Microelectronics, Zhengzhou University, Zhengzhou, Henan 450001,
China}

\begin{abstract}

In the chiral unitary approach, we have studied the single Cabibbo-suppressed decays $\Lambda_c\to
pK^+K^-$ and $\Lambda_c \to p \pi^+\pi^-$ by taking into account the
$s$-wave meson-meson interaction as well as the contributions from the intermediate vectors $\phi$ and
$\rho^0$. Our theoretical results for the ratios of the branching
fractions of $\Lambda_c\to p \bar{K}^{*0}$ and $\Lambda_c\to p
\omega$ with respect to the one of $\Lambda_c\to p \phi$ are in
agreement with the experimental data. Within the picture that the
scalar resonances $f_0(500)$, $f_0(980)$, and $a_0(980)$ are dynamically
generated from the pseudoscalar-pseudoscalar interactions in $s$-wave, we have
calculated the $K^+K^-$ and $\pi^+\pi^-$ invariant mass distributions
respectively for the decays $\Lambda_c\to pK^+K^-$ and $\Lambda_c\to
p\pi^+\pi^-$. One can find a broad bump structure for the $f_0(500)$
and a narrow peak for the $f_0(980)$ in the $\pi^+\pi^-$ invariant mass
distribution of the decay $\Lambda_c\to p\pi^+\pi^-$.
For the $K^+K^-$ invariant mass
distribution, in addition to the narrow peak for the
$\phi$ meson, there is an enhancement structure near the $K^+K^-$
threshold mainly due to the contribution from the $f_0(980)$. Both the $\pi^+\pi^-$ and $K^+K^-$ invariant mass distributions are compatible with the BESIII measurement.
We encourage our experimental colleagues to measure these
two decays, which would be helpful to understand the nature of the
$f_0(500)$, $f_0(980)$, and $a_0(980)$.
\end{abstract}



\maketitle

\section{Introduction}
\label{sec:intro}

The non-leptonic decays of the lightest charmed baryon $\Lambda_c$
play an important role in the study of strong and weak
interactions~\cite{Cheng:2015iom,Ebert:1983ih,Cheng:2018hwl,Cheng:1991sn,
Lu:2016ogy,Geng:2018upx}. In the last decades, lots of the information about the $\Lambda_c$
decays has been accumulated~\cite{Abe:2001mb,Ablikim:2015flg,Pal:2017ypp,Zupanc:2013iki,
Aaij:2017rin}, which provides a good platform to investigate the
possible final state interference effects where some resonances can
be dynamically
generated~\cite{Miyahara:2015cja,Dai:2018hqb,Xie:2018gbi,Xie:2017erh,Xie:2017xwx,Xie:2016evi}.

Recently, the BESIII Collaboration has reported the branching fractions of the $\Lambda_c\to p K^+K^-, \,p\pi^+\pi^-$,
\begin{eqnarray}
\frac{\mathcal{B}(\Lambda_c \to p \phi)}{\mathcal{B}(\Lambda_c \to p K^-\pi^+)}&=&(1.81\pm 0.33\pm 0.13)\%, \\
\frac{\mathcal{B}(\Lambda_c \to p K^+K^-)_{{\rm non}-\phi}}{\mathcal{B}(\Lambda_c \to p K^-\pi^+)}&=&(9.36\pm 2.22\pm 0.71)\%,\\
\frac{\mathcal{B}(\Lambda_c \to p \pi^+\pi^-)}{\mathcal{B}(\Lambda_c \to p K^-\pi^+)}&=&(6.70\pm 0.48\pm 0.25)\%,
\end{eqnarray}
and also measured the $\pi^+\pi^-$ and $K^+K^-$ invariant mass distributions,
respectively~\cite{Ablikim:2016tze}, where one can find a broad bump
around 500~MeV for the scalar resonance $f_0(500)$ and a narrow
 peak around 980~MeV for the scalar resonance $f_0(980)$ in the
$\pi^+\pi^-$ invariant mass distribution, in addition to the peak for the
$\rho^0$ meson. Later, the LHCb Collaboration has also reported these
ratios using the proton-proton collision data~\cite{Aaij:2017rin},
\begin{eqnarray}
\frac{\mathcal{B}(\Lambda_c \to p K^+K^-)}{\mathcal{B}(\Lambda_c \to p K^-\pi^+)}&=&(1.70\pm 0.03\pm 0.03)\%, \\
\frac{\mathcal{B}(\Lambda_c \to p \pi^+\pi^-)}{\mathcal{B}(\Lambda_c \to p K^-\pi^+)}&=&(7.44\pm 0.08\pm 0.18)\%.
\end{eqnarray}
Before the BESIII and LHCb results, the above two decay modes have
also been observed by the NA32~\cite{Barlag:1990yv},
E687~\cite{Frabetti:1993ew}, CLEO~\cite{Alexander:1995hd}, and
Belle Collaborations~\cite{Abe:2001mb}.

Within the chiral unitary approach, the scalar resonances
$f_0(500)$, $f_0(980)$, $a_0(980)$, and $K^*_0(700)$ [known as $\kappa(800)$] appear as
composite states of meson-meson, automatically dynamically generated
by the interaction of pseudoscalar-pseudoscalar where the kernel for
the Bethe-Salpter equation is taken from the chiral
Lagrangians~\cite{Oller:1997ti,Oller:1998hw,
Kaiser:1998fi,Locher:1997gr,Nieves:1999bx, Pelaez:2006nj}. The
productions of $f_0(500)$, $f_0(980)$, and $a_0(980)$ have been recently studied
with the chiral unitary approach and the final state interactions in
the decays of the $D^0$~\cite{Xie:2014tma}, $D^+_s$~\cite{Molina:2019udw}, $\bar{B}$ and
$\bar{B}_s$~\cite{Liang:2014ama,Liang:2014tia,Liang:2015qva,Xie:2018rqv},
$\chi_{c1}$~\cite{Liang:2016hmr,Kornicer:2016axs},
 $\tau^-$~\cite{Dai:2018rra}, and
$J/\psi$~\cite{Liang:2019jtr}.

In this work, we perform the calculations for the decays $\Lambda_c
\to p K^+K^-$ and $\Lambda_c \to p \pi^+\pi^-$ taking into account the meson-meson
interaction in coupled channels and also the contributions from the intermediate vector mesons $\phi$ and $\rho^0$. The  final states interaction of the pseudoscalar-pseudoscalar
in the decay $\Lambda_c \to p \pi^+\pi^-$ can propagate in $s$-wave,
which will generate the $f_0(500)$ and $f_0(980)$ resonances, and
for the decay $\Lambda_c \to p K^+K^-$, the $f_0(980)$ and $a_0(980)$ resonances
dynamically generated from the $s$-wave final state
interaction will result in an enhancement structure close to the
$K^+K^-$ threshold.

The paper is organized as follows. In Section II, we present the
formalism and ingredients for the decays of the $\Lambda_c
\to p K^+ K^-$ and $p\pi^+ \pi^-$ decays. Numerical results for the $K^+K^-$ and $\pi^+\pi^-$
invariant mass distributions   and
discussions are given in Section III, followed by a short summary in
the last section.

\section{Formalism}  \label{sec:form}

In this section, we will present the formalism for the decays
$\Lambda_c \to p K^+ K^-$ and $\Lambda_c \to p \pi^+ \pi^-$.
For the three-body decays of $\Lambda_c$, the
$s$-wave final state interactions of $\pi^+\pi^-$ or $K^+K^-$ will
dynamically generate the scalar resonances $f_0(500)$,
$f_0(980)$, and $a_0(980)$. In addition, the three-body decays can happen via the
intermediate vector mesons $\rho^0$ or $\phi$. We first introduce
the formalism for the mechanism of final state interactions of
$\pi^+\pi^-$ or $K^+K^-$ in $s$-wave in Subsect.~\ref{sec:f980},
then we show the details for the mechanism of the $\Lambda_c$ decays
via the intermediate vector mesons $\rho^0$ and $\phi$ in
Subsect~\ref{sec:vector}.

\subsection{$s$-wave final state interactions of $K^+ K^-$ and $\pi^+\pi^-$}
\label{sec:f980}

\begin{figure*}[tbhp]
\begin{center}
\includegraphics[scale=0.6]{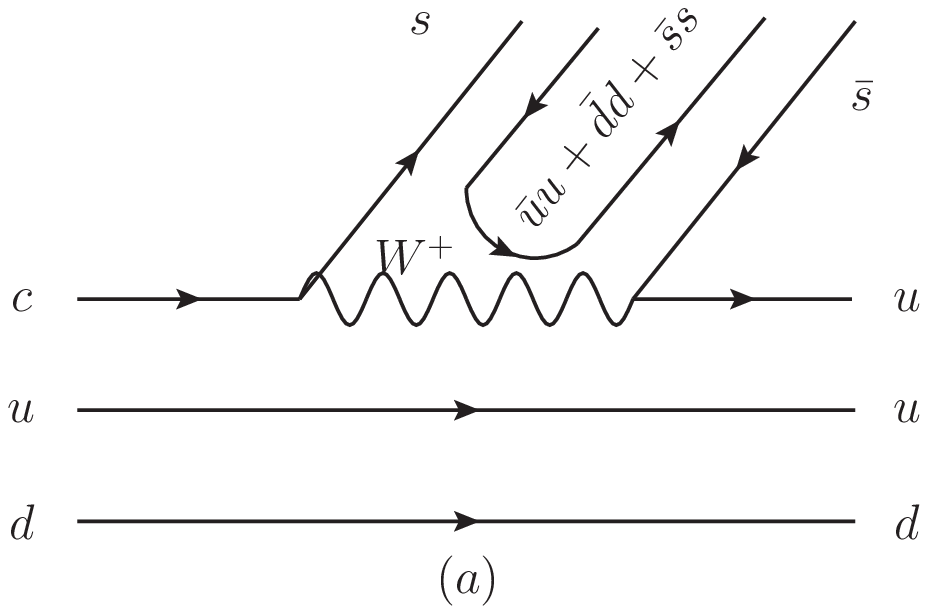}
\includegraphics[scale=0.6]{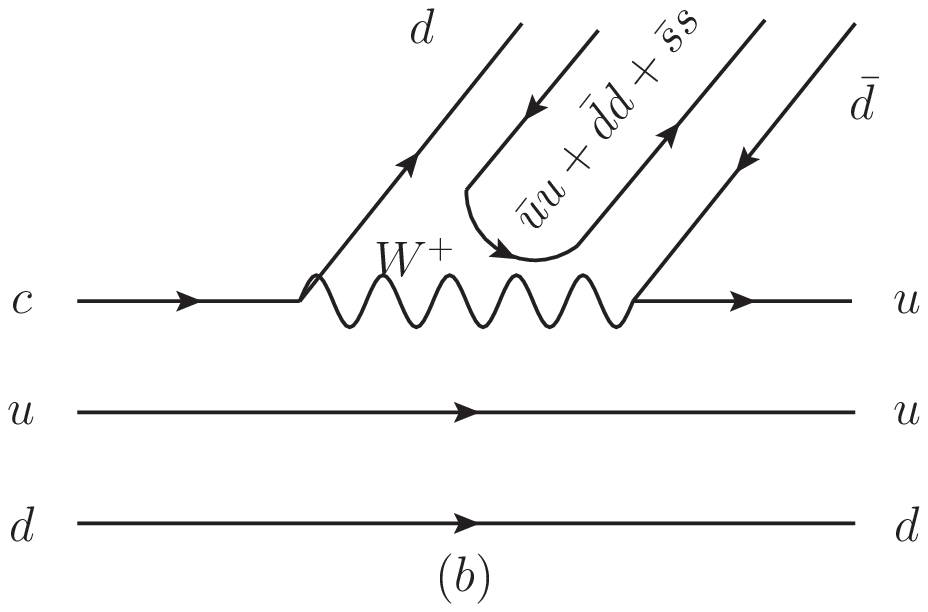}
\includegraphics[scale=0.6]{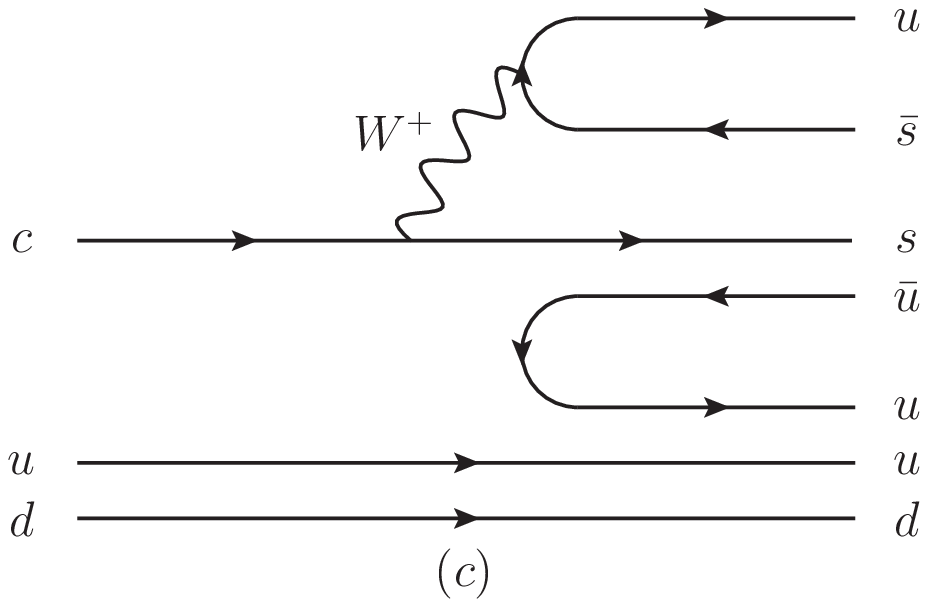}
\includegraphics[scale=0.6]{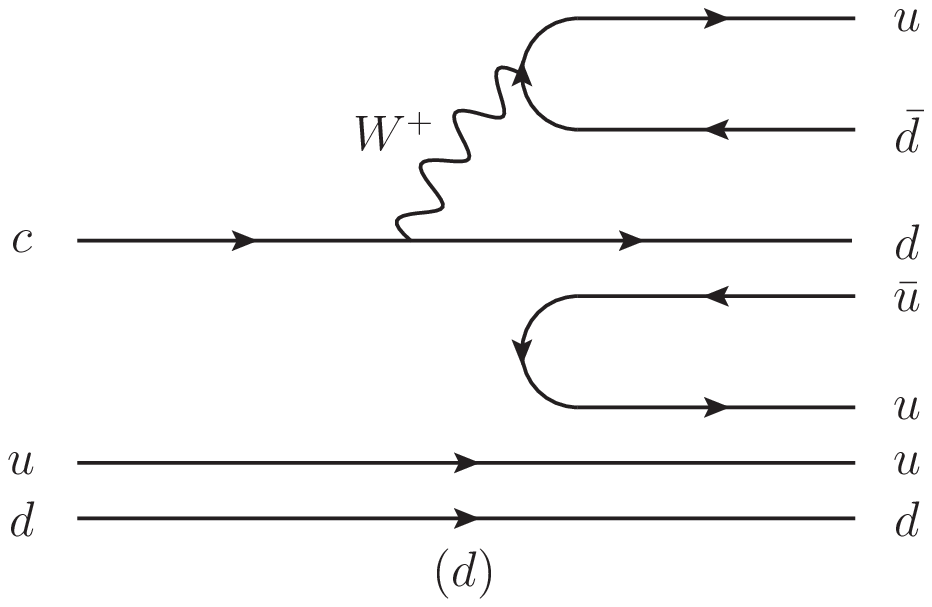}
\end{center}
\caption{The diagrams of the decays $\Lambda_c\to pK^+K^-$ and $p\pi^+\pi^-$, (a) the internal $W$ emission for $\Lambda_c \to p K^+K^-$, (b) the internal $W$ emission for $\Lambda_c \to
p \pi^+\pi^-$, (c) the external $W$ emission for $\Lambda_c \to p K^+K^-$, and (d) the external $W$ emission for $\Lambda_c \to p \pi^+\pi^-$.} \label{Fig:quarkthreebody}
\end{figure*}

Following Refs.~\cite{Oset:2016lyh,Wang:2015pcn,Lu:2016roh,Dai:2018nmw}, we take the
decay mechanism of the internal $W$ emission mechanism for the
decays $\Lambda_c \to p K^+K^-$ and $\Lambda_c \to p \pi^+ \pi^-$ as
depicted in Figs.~\ref{Fig:quarkthreebody}(a) and (b).  For the weak
decays of $\Lambda_c$, the $c$ quark decays into a $W^+$ boson and an
$s$ (or $d$) quark, then the $W^+$ boson decays into an $\bar{s}u$
(or $\bar{d}u$) pair. In order to give rise to the final states of
$p K^+K^-$ (or  $p \pi^+\pi^-$), the $s\bar{s}$ (or $d\bar{d}$)
quark pair need to hadronize together with the $\bar{q}q$
($=\bar{u}u+\bar{d}d+\bar{s}s$) produced in the vacuum, which are given by,
\begin{eqnarray}
H^{(a)}&=& V^{(a)} s(\bar{u}u+\bar{d}d+\bar{s}s)\bar{s} u\frac{1}{\sqrt{2}}(ud-du) = V^{(a)} \left(M^2\right)_{33}p, \\
H^{(b)}&=& V^{(b)} d(\bar{u}u+\bar{d}d+\bar{s}s)\bar{d}
u\frac{1}{\sqrt{2}}(ud-du) = V^{(b)}\left(M^2\right)_{22}p,
\end{eqnarray}
where $V^{(a)}$ and $V^{(b)}$ are the weak interaction strengths. We use
$\left|p\right\rangle=\frac{1}{\sqrt{2}}\left|u(ud-du)\right\rangle$,
and $\left|
\Lambda_c\right\rangle=\frac{1}{\sqrt{2}}\left|c(ud-du)\right\rangle$.
 $M$ is the $q\bar{q}$ matrix,
\begin{align*}
M=\left(\begin{array}{ccc}
              u\bar{u} & u\bar{d} & u\bar{s}\\
              d\bar{u} & d\bar{d} & d\bar{s} \\
              s\bar{u} & s\bar{d} & s\bar{s}
      \end{array}
\right)\,.
\label{eq:Mmatrix}
\end{align*}
The matrix $M$ in terms of pseudoscalar mesons can be written as,

\begin{equation}
M\Rightarrow P=
\left(\begin{array}{ccc}
              \frac{\pi^0}{\sqrt{2}}  + \frac{\eta}{\sqrt{3}}+\frac{\eta'}{\sqrt{6}}& \pi^+ & K^+\\
              \pi^-& -\frac{1}{\sqrt{2}}\pi^0 + \frac{\eta}{\sqrt{3}}+ \frac{\eta'}{\sqrt{6}}& K^0\\
               K^-& \bar{K}^0 & -\frac{\eta}{\sqrt{3}}+ \frac{2\eta'}{\sqrt{6}}
      \end{array}
\right)\,.
\label{eq:m2}
\end{equation}

Then, we have,
\begin{eqnarray}
H^{(a)} &=& V^{(a)} \left(M^2\right)_{33} p = V_P V_{cs}V_{us} \left(K^-K^+ +K^0\bar{K}^0 +\frac{1}{3}\eta\eta\right) p, \label{eq:hadA}
\\
H^{(b)} &=& V^{(b)} \left(M^2\right)_{22} p = V_P V_{cd}V_{ud}
 \left(\pi^+\pi^-+\frac{1}{2}\pi^0\pi^0+\frac{1}{3}\eta\eta-\frac{2}{\sqrt{6}}\pi^0\eta+K^0\bar{K}^0\right)p, \label{eq:hadB}
\end{eqnarray}
where we neglect the $\eta'$ because of its large mass. $V_P$ is the strength of the
 production vertex, and contains all dynamical factors, which is assumed to be same for Figs.~\ref{Fig:quarkthreebody}(a) and \ref{Fig:quarkthreebody}(b) within the SU(3) flavor symmetry. In the following, we will see that this hypothesis is also reasonable by comparing the predicted ratios of the two body decays of $\Lambda_c$ with the experimental measurements. In this work we take
$V_{cd}=V_{us}= -0.22534$, $V_{cs}=V_{ud}=0.97427$~\cite{PDG2018}.

On the other hand, the decays $\Lambda_c \to p K^+ K^-$ and
$\Lambda_c \to p \pi^+ \pi^-$ can also proceed with the color favored external $W$ emission mechanism:
i) the charmed quark turns into $W^+$ and the $s$ (or $d$)
quark, with the $K^+$ or $\pi^+$ emission from the $W^+$; ii) the
remaining quarks $s$ (or $d$) and $ud$ in the $\Lambda_c$, together with the $u\bar{u}$ pair created from the vacuum, hadronize to
the $K^- p$ (or $\pi^- p$), as depicted in Figs.~\ref{Fig:quarkthreebody}(c) and (d) respectively.
Thus, we have,
\begin{eqnarray}
H^{(c)}&=&V^{\rm (c)}(u\bar{s})s\bar{u}u\frac{1}{\sqrt{2}}(ud-du)=C\times V_P V_{cs}V_{us} K^+K^-p, \\
H^{(d)}&=&V^{\rm (d)}(u\bar{d})d\bar{u}u\frac{1}{\sqrt{2}}(ud-du)= C\times V_P V_{cd}V_{ud}\pi^+\pi^-p,
\end{eqnarray}
where we take the same normalization factor $V_P$ as Eqs.~(\ref{eq:hadA}) and (\ref{eq:hadB}),
the color factor $C$ accounts for the relative weight of the external emission mechanism  with respect to the one of the internal emission mechanism~\cite{Dai:2018nmw,Zhang:2020rqr}. The value of $C$ should be around 3,  because the  quarks from the $W$ decay in the external emission diagram [for example, the $u$ and $\bar{s}$ of Fig.~\ref{Fig:quarkthreebody}(c)] have three choices of the colors (we take $N_c=3$), while the quarks from the $W$ decay in the external emission diagram [for example, the $u$ and $\bar{s}$ of Fig.~\ref{Fig:quarkthreebody}(a)] have the fixed colors. We will keep this factor in the following formalism,  and present our results by varying its value in Sect.~\ref{sec:results}.

\begin{figure}[tbhp]
\begin{center}
\includegraphics[scale=0.6]{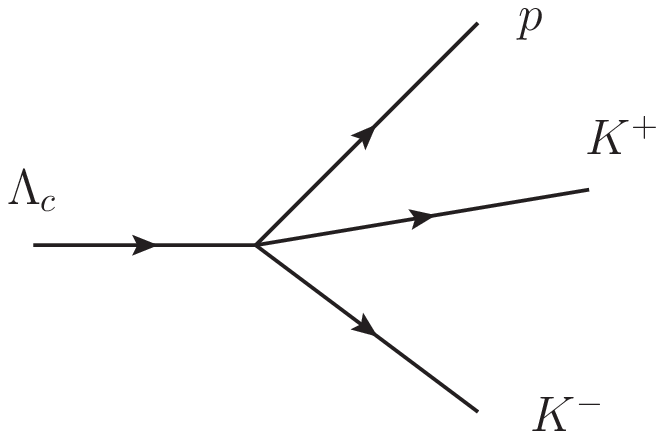}
\includegraphics[scale=0.6]{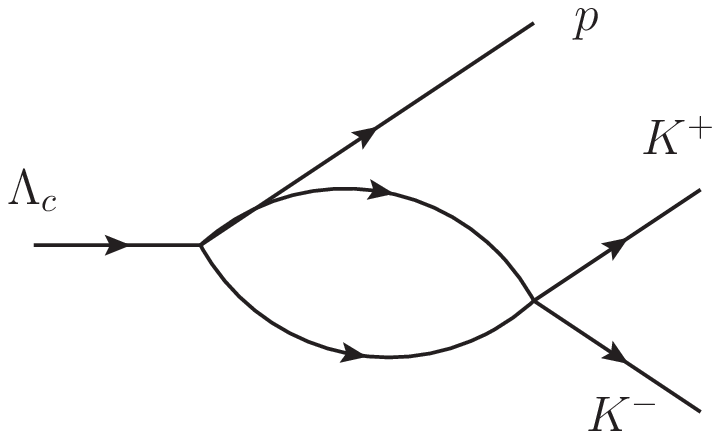}
\end{center}
\caption{The mechanisms of the decay $\Lambda_c \to p K^+K^-$, left)
tree diagram, right) the $s$-wave  final state interactions.}
\label{Fig:KKFSI}
\end{figure}

\begin{figure}[tbhp]
\begin{center}
\includegraphics[scale=0.6]{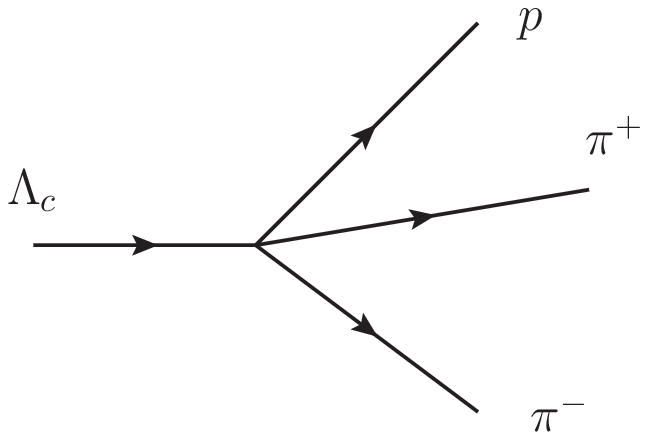}
\includegraphics[scale=0.6]{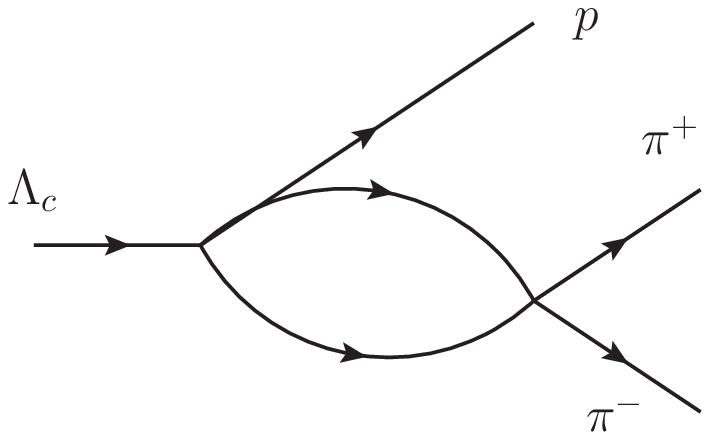}
\end{center}
\caption{The mechanisms of the decay $\Lambda_c \to p \pi^+\pi^-$,
left) tree diagram, right) the $s$-wave  final state interactions.}
\label{Fig:pipiFSI}
\end{figure}

After the production of a meson-meson pair, the final state
interaction in the $s$-wave between the mesons takes place, as shown in Figs.~\ref{Fig:KKFSI} and \ref{Fig:pipiFSI} for the decays $\Lambda_c\to p K^+K^-$ and $\Lambda_c\to p \pi^+\pi^-$.  Since the isospin of $\pi^+\pi^-$ system is $I=0$, we will take into account the  contributions from all the mechanisms of Fig.~\ref{Fig:quarkthreebody}, except the $\pi^0\eta p$ states of Eq.~(\ref{eq:hadB}) because of the isospin violation, and the amplitude is given by,
\begin{eqnarray}
t^{s-{\rm wave}}_{\Lambda_c\to p\pi^+\pi^-}&=& t^{(a)}_{\Lambda_c\to p\pi^+\pi^-}+t^{(b)}_{\Lambda_c\to p\pi^+\pi^-}+t^{(c)}_{\Lambda_c\to p\pi^+\pi^-}+t^{(d)}_{\Lambda_c\to p\pi^+\pi^-} \nonumber \\
&=&  V_PV_{cs}V_{us} \left[ G_{K^+K^-}t_{K^+K^-\to \pi^+\pi^-} +  G_{K^0\bar{K}^0}t_{K^0\bar{K}^0\to \pi^+\pi^-}   +\frac{1}{3}\times 2\times \frac{1}{2} G_{\eta\eta}\tilde{t}_{\eta\eta\to \pi^+\pi^-} \right ] \nonumber \\
&&+ V_PV_{cs}V_{us} \left[1 +G_{\pi^+\pi^-}t_{\pi^+\pi^-\to \pi^+\pi^-} + \frac{1}{2}\times 2\times \frac{1}{2} G_{\pi^0\pi^0}\tilde{t}_{\pi^0\pi^0\to \pi^+\pi^-}   \right. \nonumber\\
&&\left. +\frac{1}{3}\times 2 \times \frac{1}{2}
G_{\eta\eta}\tilde{t}_{\eta\eta\to \pi^+\pi^-} +  G_{K^0\bar{K}^0}t_{K^0\bar{K}^0\to \pi^+\pi^-}  \right] \nonumber \\
&&+ C\times V_PV_{cs}V_{us} \left[G_{K^+K^-}t_{K^+K^-\to \pi^+\pi^-}\right] + C\times V_P \left[1+ G_{\pi^+\pi^-}t_{\pi^+\pi^-\to \pi^+\pi^-}\right] \nonumber \\
&=& V_P V_{cs}V_{us} \left[(1+C) +(1+C) G_{K^+K^-}t_{K^+K^-\to \pi^+\pi^-} + 2 G_{K^0\bar{K}^0}t_{K^0\bar{K}^0\to \pi^+\pi^-} + (1+C) G_{\pi^+\pi^-}t_{\pi^+\pi^-\to \pi^+\pi^-} \right. \nonumber \\
&& \left. + \frac{1}{2} G_{\pi^0\pi^0}\tilde{t}_{\pi^0\pi^0\to
\pi^+\pi^-}  +\frac{2}{3}G_{\eta\eta}\tilde{t}_{\eta\eta\to \pi^+\pi^-} \right ].
\label{eq:amp_f980_pipi}
\end{eqnarray}

For the decay $\Lambda_c\to p K^+K^-$, the amplitude is given by,
\begin{eqnarray}
t^{s-{\rm wave}}_{\Lambda_c\to pK^+K^-}&=& t^{(a)}_{\Lambda_c\to pK^+K^-}+ t^{(b)}_{\Lambda_c\to pK^+K^-}+ t^{(c)}_{\Lambda_c\to pK^+K^-}+ t^{(d)}_{\Lambda_c\to pK^+K^-} \nonumber \\
&=& V_PV_{cs}V_{us}\left[1+  G_{K^+K^-}t_{K^+K^-\to K^+K^-} +  G_{K^0\bar{K}^0}t_{K^0\bar{K}^0\to K^+K^-}  +\frac{1}{3}\times 2\times \frac{1}{2} G_{\eta\eta}\tilde{t}_{\eta\eta\to K^+K^-} \right ] \nonumber \\
&&+ V_P V_{cs}V_{us}\left[G_{\pi^+\pi^-}t_{\pi^+\pi^-\to K^+K^-} + \frac{1}{2}\times 2\times \frac{1}{2} G_{\pi^0\pi^0}\tilde{t}_{\pi^0\pi^0\to  K^+K^-}  +\frac{1}{3}\times 2 \times \frac{1}{2}
G_{\eta\eta}\tilde{t}_{\eta\eta\to K^+K^-} \right. \nonumber \\
&& \left. - \frac{2}{\sqrt{6}} G_{\pi^0\eta}t_{\pi^0\eta\to
K^+K^-} +  G_{K^0\bar{K}^0}t_{K^0\bar{K}^0\to K^+K^-}  \right] \nonumber \\
&&+ C\times V_P V_{cs}V_{us}\left[1+G_{K^+K^-}t_{K^+K^-\to K^+K^-}\right] + C\times V_P \left[ G_{\pi^+\pi^-}t_{\pi^+\pi^-\to K^+K^-}\right] \nonumber \\
&=& V_PV_{cs}V_{us} \left[ (1+C)+ G_{K^+K^-}t_{K^+K^-\to K^+K^-} +  G_{K^0\bar{K}^0}t_{K^0\bar{K}^0\to K^+K^-} + (1+C) G_{\pi^+\pi^-}t_{\pi^+\pi^-\to K^+K^-} \right. \nonumber \\
&&\left. +\frac{1}{2} G_{\pi^0\pi^0}\tilde{t}_{\pi^0\pi^0\to
K^+K^-} + \frac{2}{3}G_{\eta\eta}\tilde{t}_{\eta\eta\to
K^+K^-}\right ] \nonumber \\
&& + V_PV_{cs}V_{us}\left[ C\times G_{K^+K^-}t_{K^+K^-\to K^+K^-} + G_{K^0\bar{K}^0}t_{K^0\bar{K}^0\to K^+K^-} - \frac{2}{\sqrt{6}} G_{\pi^0\eta}t_{\pi^0\eta\to
K^+K^-} \right], \label{eq:amp_f980_KK}
\end{eqnarray}
where the first term only contains the contribution from isospin $I=0$, and the second term has the contributions of  $I=0$ and  $I=1$ from the mechanisms of Figs.~\ref{Fig:quarkthreebody}(b) and (c). It is easily done taking  $G_{K^0\bar{K}^0}=G_{K^+K^-}$, and rewriting $t_{K^+K^-\to K^+K^-}$ and $t_{K^0\bar{K}^0\to K^+K^-}$ from the mechanisms of Figs.~\ref{Fig:quarkthreebody}(b) and (c) as in Ref.~\cite{Liang:2015qva},
\begin{eqnarray}
&&  G_{K^0\bar{K}^0}t_{K^0\bar{K}^0\to K^+K^-}+ C\times G_{K^+K^-}t_{K^+K^-\to K^+K^-}  \nonumber \\&=& G_{K^0\bar{K}^0}\left[ \frac{1+C}{2} \left(t_{K^0\bar{K}^0\to K^+K^-} +t_{K^+K^-\to K^+K^-} \right) +\frac{1-C}{2} \left( t_{K^0\bar{K}^0\to K^+K^-} -t_{K^+K^-\to K^+K^-}  \right) \right],
\end{eqnarray}
where the first two terms are in $I=0$ while the last two terms are in $I=1$.

Thus, the amplitude of Eq.~(\ref{eq:amp_f980_KK}) can be rewritten as,
\begin{eqnarray}
t^{s-{\rm wave}}_{\Lambda_c\to pK^+K^-}
&=& V_P V_{cs}V_{us}\left\lbrace \left[ (1+C) + \frac{3+C}{2}G_{K^0\bar{K}^0} \left(t_{K^0\bar{K}^0\to K^+K^-} +t_{K^+K^-\to K^+K^-} \right) \right. \right. \nonumber \\
&&\left. + (1+C) G_{\pi^+\pi^-}t_{\pi^+\pi^-\to K^+K^-} +\frac{1}{2} G_{\pi^0\pi^0}\tilde{t}_{\pi^0\pi^0\to
K^+K^-} +\frac{2}{3}G_{\eta\eta}\tilde{t}_{\eta\eta\to
K^+K^-}  \right ] \nonumber \\
&&\left. +\left[  \frac{1-C}{2} G_{K^0\bar{K}^0}\left( t_{K^0\bar{K}^0\to K^+K^-} -t_{K^+K^-\to K^+K^-}\right)  - \frac{2}{\sqrt{6}} G_{\pi^0\eta}t_{\pi^0\eta\to
K^+K^-} \right] \right\rbrace \nonumber \\
&=& t^{I=0} +  t^{I=1}  , \label{eq:amp_f980_KK2}
\end{eqnarray}
where the terms $t^{I=0}$ and $t^{I=1}$ correspond to the contributions from the $I=0$ and $I=1$, respectively.

In Eqs.~(\ref{eq:amp_f980_pipi}) and (\ref{eq:amp_f980_KK}), we include a factor of 2 from the two way to match the two identical particles of the operators in Eqs.~(\ref{eq:hadA}) and (\ref{eq:hadB}) with the two mesons ($\pi^0\pi^0$ and $\eta\eta$) produced, and  a factor $1/2$ in the intermediate loops involving a pair of identical mesons~\cite{Liang:2014ama,Liang:2015qva}. The scattering matrix $t_{i\to j}$ has been calculated within the chiral unitary approach in Refs.~\cite{Oller:1997ti,Guo:2005wp,Liang:2014tia,Xie:2014tma,Dias:2016gou}, and we take $\tilde{t}_{\eta\eta\to i}=\sqrt{2}t_{\eta\eta\to i}$, $\tilde{t}_{\pi^0\pi^0\to j}=\sqrt{2} t_{\pi^0\pi^0\to j}$ for the  two identical  particles~\cite{Dias:2016gou}.
$G_l$ is the loop function for the two mesons propagator in the
$l$th channel, which is given as follows after the integration in $dq^0$,
\begin{eqnarray}
G_l &=&i\int
\frac{d^4q}{(2\pi)^4}\frac{1}{(p-q)^2-m^2_1+i\epsilon}\frac{1}{q^2-m^2_2+i\epsilon}
= \int \frac{d^3\vec q}{(2\pi)^3}\frac{\omega_1+\omega_2}{\omega_1
\omega_2}\frac{1}{(\sqrt{s}+\omega_1+\omega_2)(\sqrt{s}-\omega_1-\omega_2+i\epsilon)},\label{eq:loop}
\end{eqnarray}
where $\sqrt{s}$ is the invariant mass of the meson-meson pair, and
the meson energies $\omega_i=\sqrt{(\vec{q}\,)^2+m_i^2}$ ($i=1,2$).
The integral on $\vec{q}$ in Eq.~(\ref{eq:loop}) is performed with a
cutoff $|\vec{q}_{\rm max}|=600$~MeV, as used in
Refs.~\cite{Dias:2016gou,Liang:2014tia,Xie:2014tma}. The transition
amplitude $t_{ij}$ is obtained by solving the Bethe-Salpeter
equation in coupled channels,
\begin{equation}
T=[1-VG]^{-1}V,
\end{equation}
where five channels $\pi^+\pi^-$ (1), $\pi^0\pi^0$  (2), $K^+K^-$  (3),
$K^0\bar{K}^0$  (4), and $\eta\eta$  (5) are included for $I=0$, and three channels $K^+K^-$ (1), $K^0\bar{K}^0$, (2) and $\pi^0\eta$ (3) are included for $I=1$. The elements of the
diagonal matrix $G$ are given by the loop function of
Eq.~(\ref{eq:loop}), and $V$ is the matrix of the interaction kernel
corresponding to the tree level transition amplitudes obtained from
phenomenological Lagrangians~\cite{Oller:1997ti}. The explicit expressions for $I=0$ can be
expressed as~\cite{Dias:2016gou},
\begin{eqnarray}
&& V_{\pi^+\pi^-\to\pi^+\pi^-}=-\frac{1}{2f^2}s,~~~V_{\pi^+\pi^-\to\pi^0\pi^0}=-\frac{1}{\sqrt{2}f^2}(s-m^2_\pi),~~~V_{\pi^+\pi^-\to K^+K^-}=-\frac{1}{4f^2}s, \nonumber \\
&&V_{\pi^+\pi^-\to K^0\bar{K}^0}=-\frac{1}{4f^2}s,~~~V_{\pi^+\pi^-\to\eta\eta}=-\frac{1}{3\sqrt{2}f^2}m^2_\pi,~~~V_{\pi^0\pi^0 \to \pi^0\pi^0}=-\frac{1}{2f^2}m^2_\pi,\nonumber \\
&& V_{\pi^0\pi^0 \to K^+K^-}=-\frac{1}{4\sqrt{2} f^2}s,~~~V_{\pi^0\pi^0 \to K^0\bar{K}^0}=-\frac{1}{4\sqrt{2} f^2}s,~~~V_{\pi^0\pi^0 \to \eta\eta}=-\frac{1}{6f^2}m^2_\pi, \nonumber \\
&& V_{K^+K^- \to K^+K^-}=-\frac{1}{2f^2}s,~~~V_{K^+K^-\to K^0\bar{K}^0}=-\frac{1}{4f^2}s,~~~V_{K^+K^-  \to \eta\eta}=-\frac{1}{12\sqrt{2}f^2}(9s-6m^2_\eta-2m^2_\pi),\nonumber \\
&& V_{K^0\bar{K}^0 \to K^0\bar{K}^0}=-\frac{1}{2f^2}s,~~~V_{K^0\bar{K}^0\to   \eta\eta}=-\frac{1}{12\sqrt{2}f^2}(9s-6m^2_\eta-2m^2_\pi),~~~V_{\eta\eta \to \eta\eta }=-\frac{1}{18f^2}(16m^2_K-7m^2_\pi),
\end{eqnarray}
and the ones for $I=1$ are~\cite{Xie:2014tma},
\begin{eqnarray}
&& V_{K^+K^- \to K^+K^-}=-\frac{1}{2f^2}s,~~~V_{K^+K^-\to K^0\bar{K}^0}=-\frac{1}{4f^2}s,~~~V_{K^+K^-\to\pi^0\eta}=\frac{-\sqrt{3}}{12f^2}\left(3s-\frac{8}{3}m^2_K-\frac{1}{3}m^2_\pi-m^2_\eta \right),\nonumber \\
&& V_{K^0\bar{K}^0-\to K^0\bar{K}^0 }=-\frac{1}{2f^2}s,~~~V_{K^0\bar{K}^0-\to \pi^0\eta}=-V_{K^+K^- \to\pi^0\eta},~~~V_{\pi^0\eta \to \pi^0\eta }=-\frac{m^2_\pi}{3f^2},
\end{eqnarray}
where $f=f_\pi=93$~MeV is the pion decay constant, and $m_\pi$, $m_K$, and $m_\eta$ are the averaged masses of the pion, kaon, and $\eta$ mesons, respectively~\cite{PDG2018}.

With the amplitudes of Eqs.~(\ref{eq:amp_f980_KK}) and (\ref{eq:amp_f980_pipi}), we can write the differential decay width for the decays $\Lambda_c\to p K^+K^-$ and $\Lambda_c\to p \pi^+\pi^-$ in $s$-wave,
\begin{equation}
\frac{d\Gamma^{s-{\rm wave}}}{dM_{\rm inv}}=\frac{1}{(2\pi)^3}\frac{p_p \tilde{k}}{4M^2_{\Lambda_c}}\left| t^{s-{\rm wave}}_{\Lambda_c\to pK^+K^-,\, p \pi^+\pi^-}\right|^2, \label{eq:dw_threebody}
\end{equation}
where $M_{\rm inv}$ is the invariant mass of the $K^+K^-$ or $\pi^+\pi^-$, $p_p$ is the momentum of the proton in the $\Lambda_c$ rest frame, and $\tilde{k}$ is the momentum of the $K^+$ (or $\pi^+$) in the rest frame of the $K^+K^-$ (or $\pi^+\pi^-$) system,
\begin{eqnarray}
p_p&=&\frac{\lambda^{1/2}\left(M^2_{\Lambda_c}, M^2_p, M^2_{\rm inv}
\right)}{2M_{\Lambda_c}}, ~~~~~~~~~~ \tilde{k} =
\frac{\lambda^{1/2}\left(M^2_{\rm inv}, m^2_{K^+/\pi^+},
m^2_{K^-/\pi^-} \right)}{2M_{\rm inv}},
\end{eqnarray}
with the K\"{a}llen function $\lambda(x,y,z)=x^2+y^2+z^2-2xy-2yz-2zx$. The masses of the baryons  and mesons involved in our calculations are taken from PDG~\cite{PDG2018}.

\subsection{$\Lambda_c$ decays via the intermediate vector mesons $\phi$ and $\rho^0$}
\label{sec:vector} 
\begin{figure}[tbhp]
\begin{center}
\includegraphics[scale=0.6]{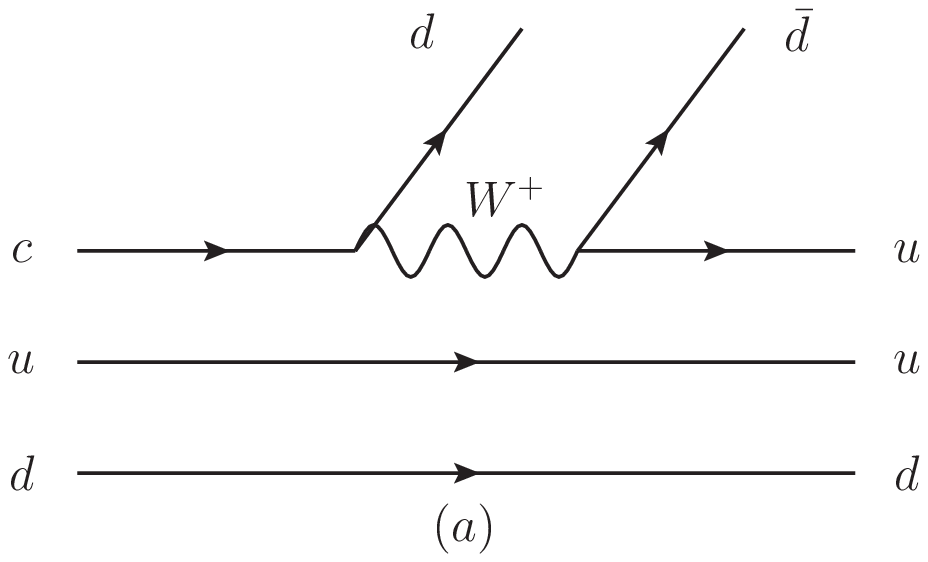}
\includegraphics[scale=0.6]{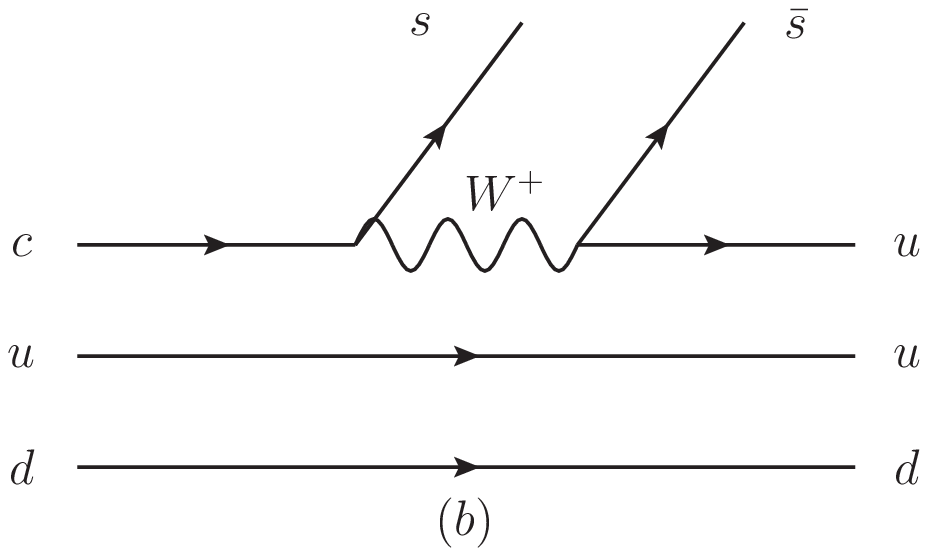}
\includegraphics[scale=0.6]{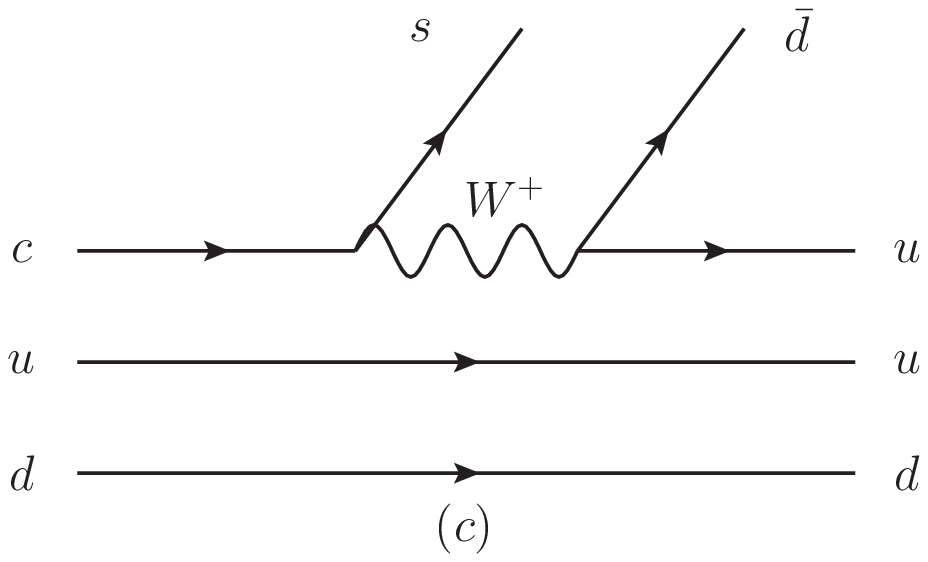}
\end{center}
\caption{The quark level diagrams for the two-body decays of
$\Lambda_c$, a) $\Lambda_c \to p \rho^0$, and $p\omega$, b)
$\Lambda_c \to p \phi$, and c)  $\Lambda_c \to p \bar{K}^{*0}$.}
\label{Fig:quarkphirho}
\end{figure}

In this section, we will present the formalism
for the decays $\Lambda_c\to p K^+K^-$ and $\Lambda_c\to p
\pi^+\pi^-$ via the intermediate mesons $\phi$ and $\rho^0$.
The quark level diagrams for the two-body decays of $\Lambda_c$ into a proton and a vector meson are shown in Fig.~\ref{Fig:quarkphirho}.

At the quark level, the quark components of the vector mesons are,
\begin{gather}
\rho^0 =\frac{1}{\sqrt{2}}(u\bar{u}-d
\bar{d}),~~~\phi=s\bar{s},~~~\omega =\frac{1}{\sqrt{2}}(u\bar{u}+d
\bar{d}), ~~\bar{K}^{*0}= s\bar{d}.
\end{gather}
The amplitudes can be written as,
\begin{eqnarray}
t_{\Lambda_c\to p\rho^0} &=&-\frac{1}{\sqrt{2}} V'_P V_{cd}V_{ud},
~~~~~~~~~
t_{\Lambda_c\to p\phi} = V'_P V_{cs}V_{us}, \label{eq:amp2body} \\
t_{\Lambda_c\to p\omega} &=&\frac{1}{\sqrt{2}} V'_P V_{cd}V_{ud},
~~~~~~~~~  t_{\Lambda_c\to p \bar{K}^{*0}} = V'_P V_{cs}V_{ud},
\end{eqnarray}
where $V'_P$ is a normalization factor for the
$\Lambda_c$ decay into proton and a vector meson. The factor of
$1/\sqrt{2}$ in the above amplitudes
comes from the quark component of the $\rho^0$ and $\omega$.  With
those amplitudes, the decay width for the two-body decay of $\Lambda_c$ into proton and a vector meson in $s$-wave
is,
\begin{equation}
\Gamma_{\Lambda_c\to p V} =
\frac{\lambda^{1/2}\left(M^2_{\Lambda_c}, m^2_V, M^2_p
\right)}{16\pi M^3_{\Lambda_c}} \left| t_{\Lambda_c\to p V}\right|^2
, \label{eq:2decay}
\end{equation}
where $V$ stands for the vector mesons $\rho^0$, $\phi$, $\omega$,
and $\bar{K}^{*0}$.

The $K^+K^-$ and  $\pi^+\pi^-$ invariant mass distributions respectively for
the $\phi$ and $\rho^0$ mesons can be obtained by converting the total
rate for vector production into a mass distribution as
Refs.~\cite{Bayar:2014qha,Liang:2014ama},
\begin{eqnarray}
\frac{d\Gamma_{\Lambda_c\to p \rho^0, \rho^0 \to \pi^+
\pi^-}}{dM_{\rm inv}} &=& \frac{2m^2_\rho}{\pi}
\frac{\tilde\Gamma_\rho \tilde{\Gamma}_{\Lambda_c\to p
\rho^0}}{(M^2_{\rm inv}-m^2_\rho)^2 + m^2_\rho
\tilde\Gamma^2_\rho } , \\
\frac{d\Gamma_{\Lambda_c\to p \phi, \phi\to K^+ K^-}}{dM_{\rm inv}}
&=& \frac{m^2_\phi}{\pi} \frac{\tilde\Gamma_\phi
\tilde{\Gamma}_{\Lambda_c\to p \phi}}{(M^2_{\rm inv}-m^2_\phi)^2 +
m^2_\phi \tilde\Gamma^2_\phi} ,
\end{eqnarray}
where we have considered that the $K^+ K^-$ decay accounts for 1/2
of the $K\bar{K}$ decay width of the $\phi$ meson. Since $\rho^0 \to
\pi^+\pi^-$ and $\phi \to K^+ K^-$ are in $p$-wave, we take
\begin{eqnarray}
\tilde\Gamma_\rho = \Gamma_{\rho^0} \left(\frac{ \sqrt{M^2_{\rm inv}
- 4m^2_\pi} }{ \sqrt{m^2_\rho - 4m^2_\pi} }  \right)^3,
 ~~~~~~~~~~~~~ \tilde\Gamma_\phi = \Gamma_{\phi} \left(\frac{ \sqrt{M^2_{\rm inv} -
4m^2_K} }{ \sqrt{m^2_\phi - 4m^2_K} }  \right)^3,
\end{eqnarray}
and
\begin{eqnarray}
\tilde\Gamma_{\Lambda_c\to p V}&=& \Gamma_{\Lambda_c\to p V}
\frac{\lambda^{1/2}\left(M^2_{\Lambda_c}, M^2_{\rm inv}, M^2_{p}
\right)}{\lambda^{1/2}\left(M^2_{\Lambda_c}, m^2_V, M^2_{p} \right)}
\frac{m_V}{M_{\rm inv}}.
\end{eqnarray}

For the processes $\Lambda_c\to p K^+K^-$ and $\Lambda_c\to p \pi^+\pi^-$, the contributions from the vector mesons $\phi$ and $\rho^0$, which respectively decay into $K^+K^-$ and $\pi^+\pi^-$ in $p$-wave, should be added to Eq.~(\ref{eq:dw_threebody}) incoherently.

\section{Results and Discussion}
\label{sec:results}

\begin{figure}[tbhp]
\begin{center}
\includegraphics[scale=0.6]{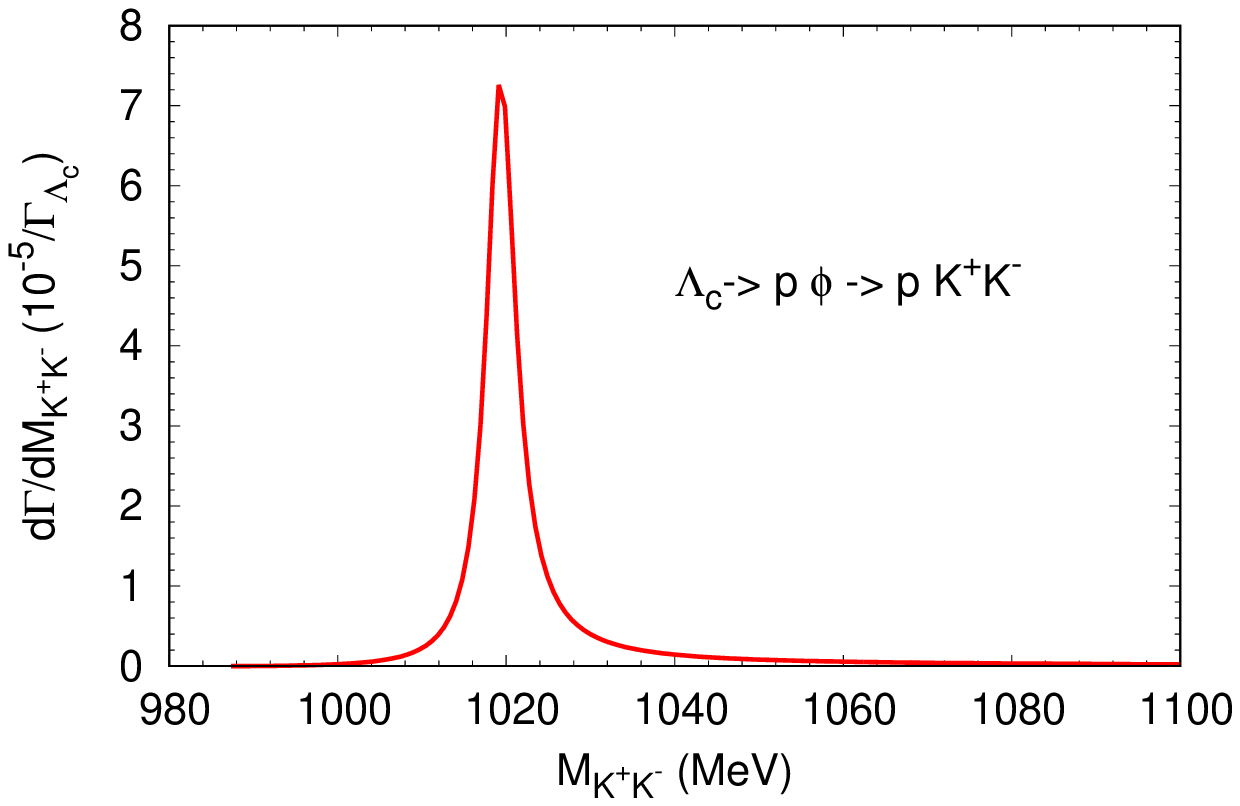}
\end{center}
\caption{The $K^+K^-$ invariant mass distribution of the decay $\Lambda_c
\to p\phi \to pK^+K^-$ with $(V'_P)^2/\Gamma_{\Lambda_c}=4.5\times
10^3$~MeV. }
\label{Fig:dw_phi}
\end{figure}

\begin{figure}[tbhp]
\begin{center}
\includegraphics[scale=0.6]{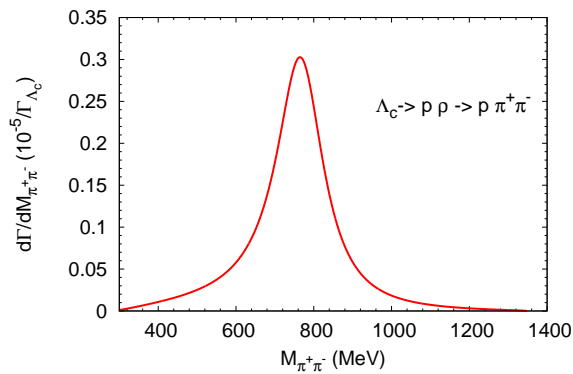}
\end{center}
\caption{The $\pi^+\pi^-$ invariant mass distribution of the decay $\Lambda_c
\to p\rho \to p\pi^+\pi^-$ with $(V'_P)^2/\Gamma_{\Lambda_c}=4.5\times
10^3$~MeV. }
\label{Fig:dw_rho}
\end{figure}

With Eqs.~(\ref{eq:amp2body}-\ref{eq:2decay}), the
ratios of the branching fractions of the decays $\Lambda_c\to p
\bar{K}^{*0}$, $\Lambda_c\to p \omega$, $\Lambda_c\to p \rho^0$ with
respect to the decay $\Lambda_c\to p \phi$ can be obtained with Eq.~(\ref{eq:2decay}),
\begin{eqnarray}
R^{\rm th}_1 &=& \frac{\mathcal{B}(\Lambda_c\to p \bar{K}^{*0})}{\mathcal{B}(\Lambda_c\to p \phi)}=\frac{\Gamma_{\Lambda_c\to p \bar{K}^{*0}}}{\Gamma_{\Lambda_c\to p \phi}} \nonumber \\
&=& \frac{\lambda^{1/2}\left(M^2_{\Lambda_c}, m^2_{\bar{K}^{*0}}, M^2_p
\right)\left| t_{\Lambda_c\to p \bar{K}^{*0}}\right|^2}{\lambda^{1/2}\left(M^2_{\Lambda_c}, m^2_{\phi}, M^2_p
\right)\left| t_{\Lambda_c\to p \phi}\right|^2}=\frac{\lambda^{1/2}\left(M^2_{\Lambda_c}, m^2_{K^{*0}}, M^2_p
\right)V^2_{ud}}{\lambda^{1/2}\left(M^2_{\Lambda_c}, m^2_{\phi}, M^2_p
\right)V^2_{us}}=21.6,  
\end{eqnarray}
and analogously,
\begin{eqnarray}
R^{\rm th}_2 &=& \frac{\mathcal{B}(\Lambda_c\to p \omega)}{\mathcal{B}(\Lambda_c\to p \phi)}=0.636, ~~~
R^{\rm th}_3 = \frac{\mathcal{B}(\Lambda_c\to p
\rho^0)}{\mathcal{B}(\Lambda_c\to p \phi)}=0.640,
\end{eqnarray}
and one can find that $R^{\rm th}_1$ and $R^{\rm th}_2$ are consistent with the experimental results~\cite{PDG2018},
\begin{eqnarray}
R^{\rm exp}_1&=&\frac{\mathcal{B}(\Lambda_c\to p \bar{K}^{*0})}{\mathcal{B}(\Lambda_c\to p \phi)}=\frac{(1.94\pm0.27) \%}{(1.06\pm0.14)\times 10^{-3}}=18.3\pm 3.5, \\
R^{\rm exp}_2&=&\frac{\mathcal{B}(\Lambda_c\to p
\omega)}{\mathcal{B}(\Lambda_c\to p \phi)}=\frac{(9\pm4)\times
10^{-4}}{(1.06\pm0.14)\times 10^{-3}}=0.85\pm0.39,
\end{eqnarray}
which implies that it is reasonable to take the same value of $V'_P$ for the mechanisms of Fig.~\ref{Fig:quarkphirho}.
By fitting to the branching fractions of the decays $\Lambda_c\to p
\bar{K}^{*0}$, $\Lambda_c\to p \phi$, and $\Lambda_c\to p \omega$,
we can obtain the $(V'_P)^2/\Gamma_{\Lambda_c}=(4.5 \pm 0.4)\times
10^3$~MeV. With this value, the branching fraction of the decay
$\Lambda_c\to p \rho^0$ is estimated to be $\mathcal{B}(\Lambda_c\to
p \rho^0)=(6.3 \pm 0.6)\times 10^{-4}$, and the $K^+K^-$ and $\pi^+\pi^-$ invariant mass distribution of the decays $\Lambda_c\to p \phi \to p K^+K^-$ and $\Lambda_c\to p \rho \to p \pi^+\pi^-$ are easily calculated as shown in Figs.~\ref{Fig:dw_phi} and \ref{Fig:dw_rho}, respectively.

\begin{figure}[tbhp]
\begin{center}
\includegraphics[scale=0.6]{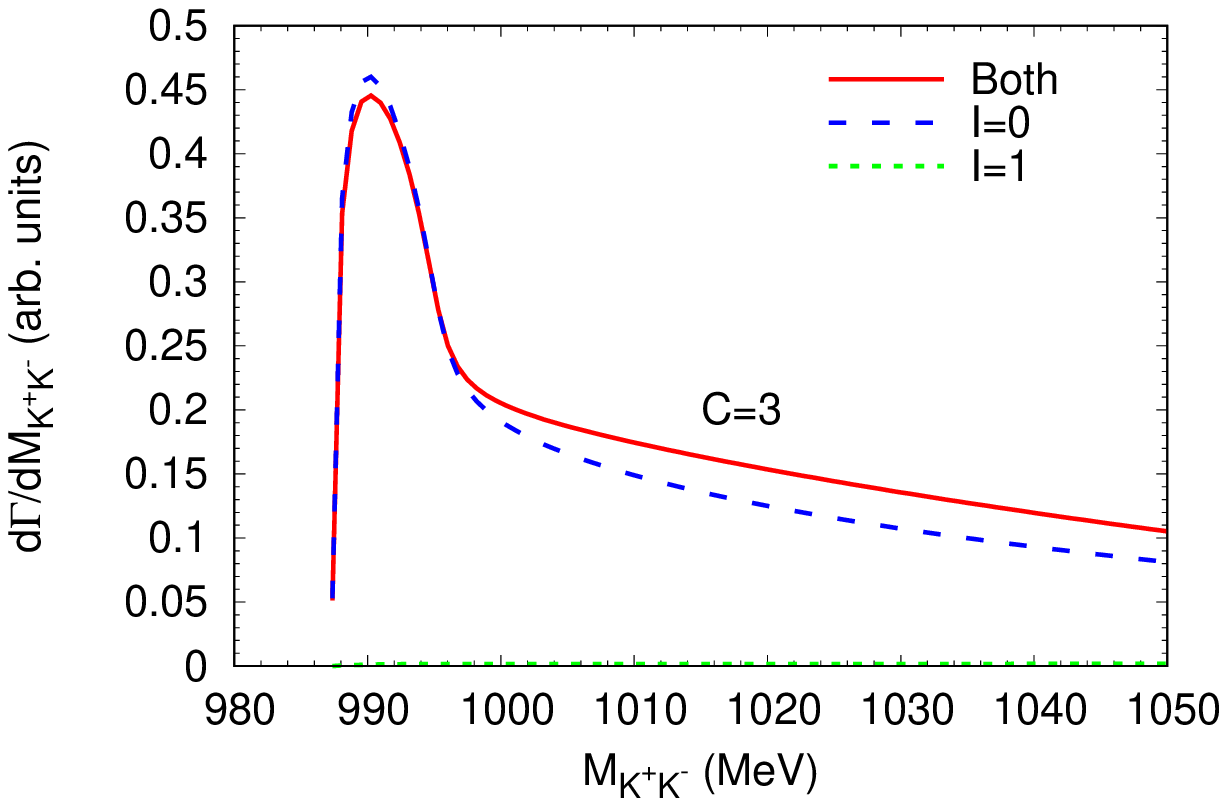}
\includegraphics[scale=0.6]{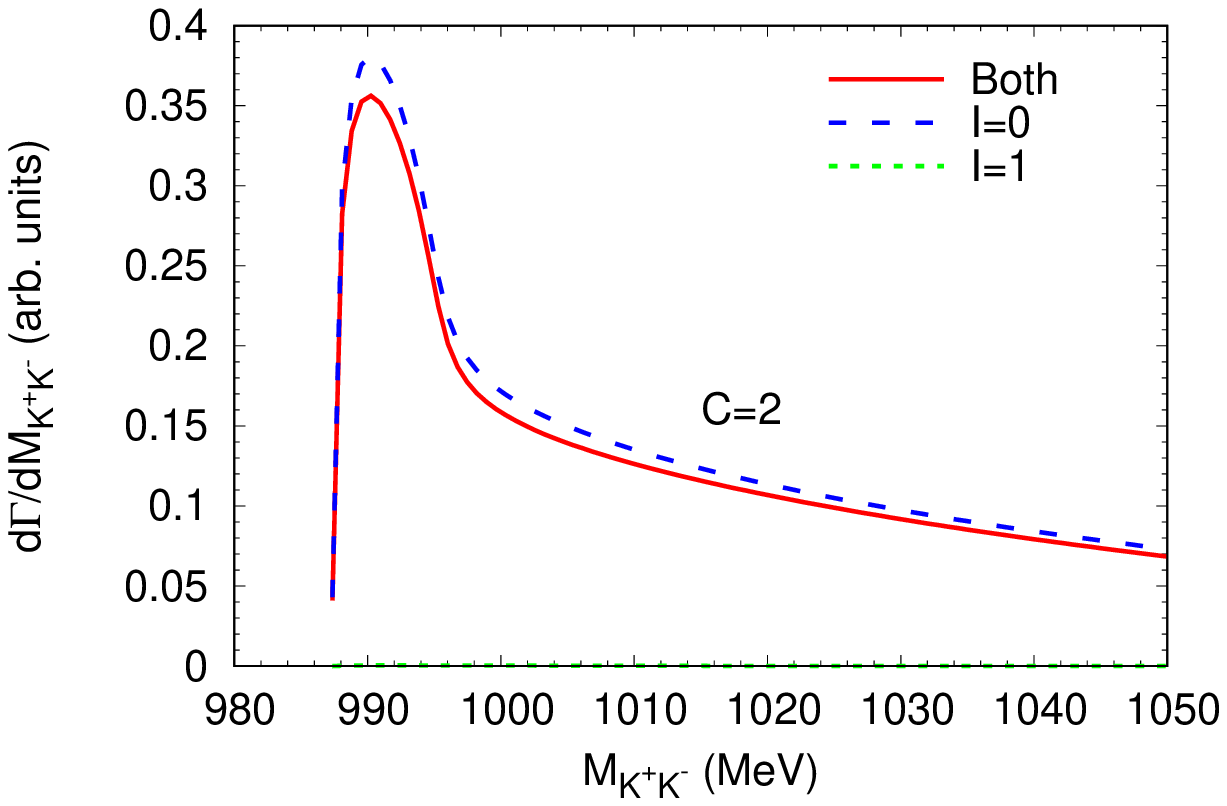}
\includegraphics[scale=0.6]{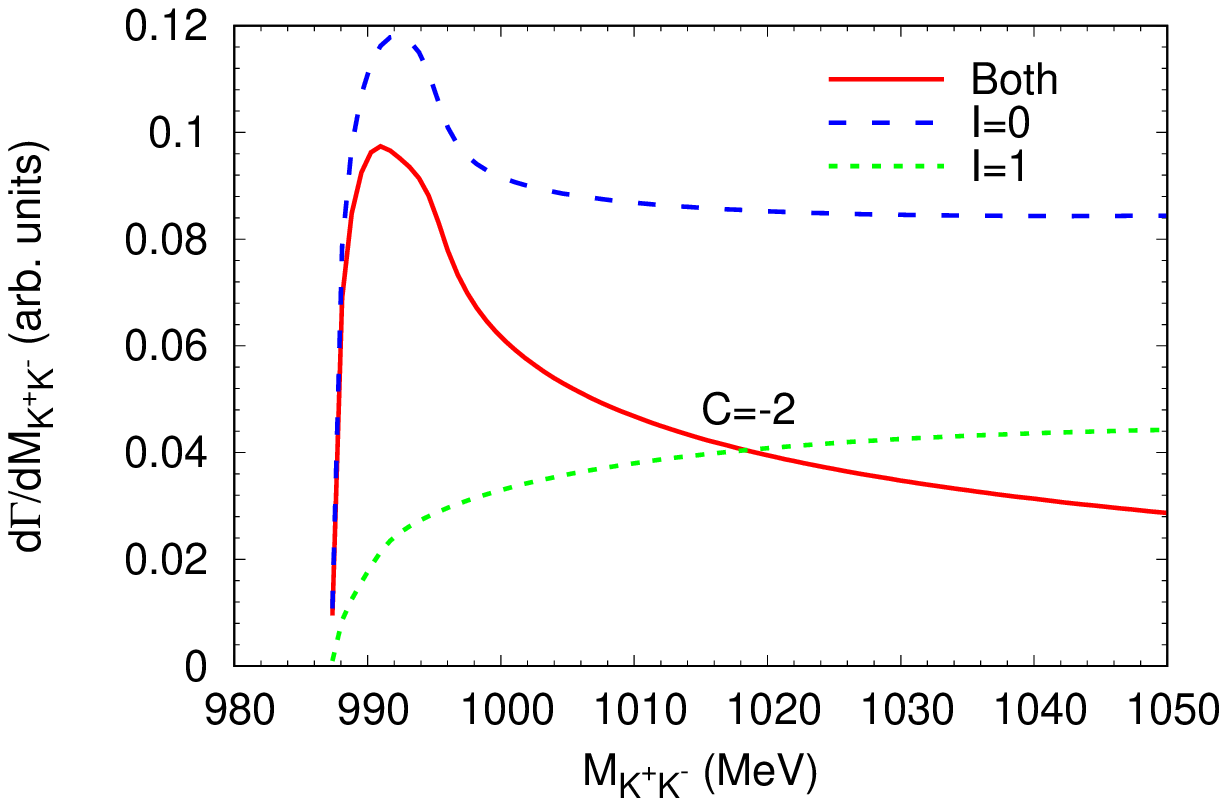}
\includegraphics[scale=0.6]{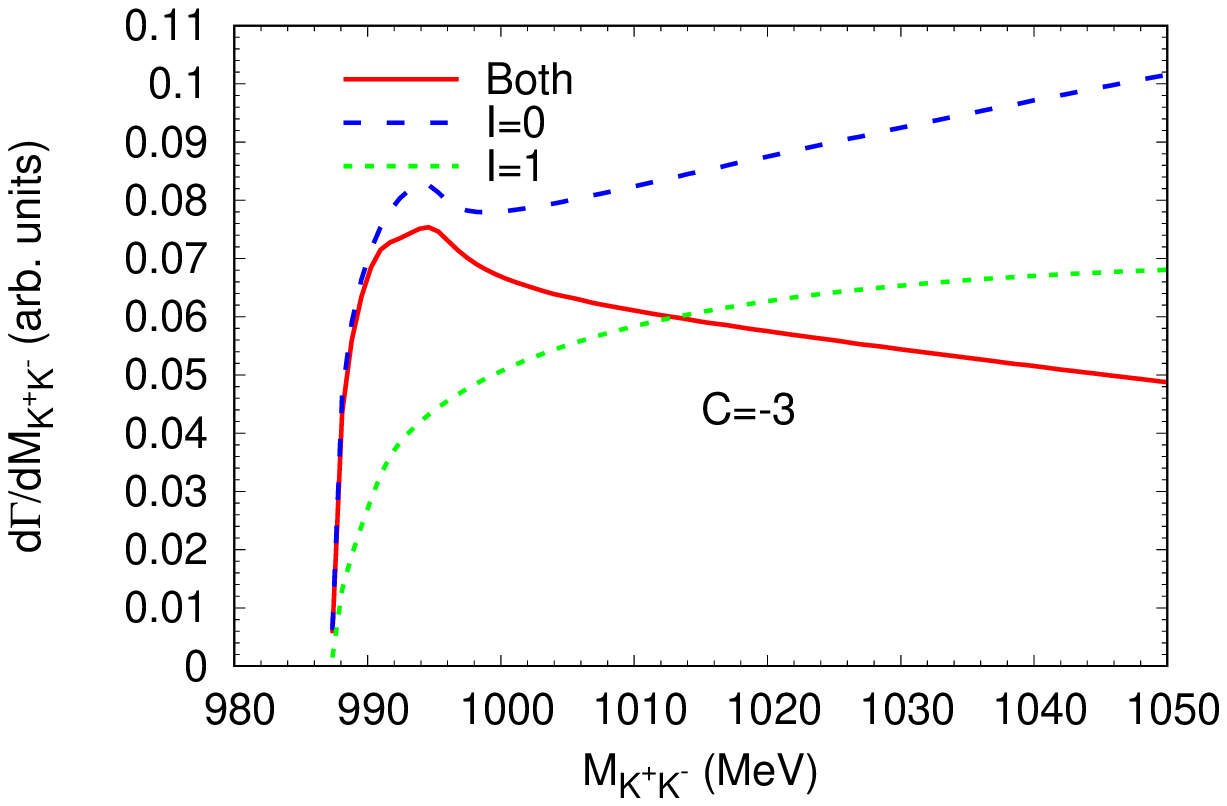}
\end{center}
\caption{The $K^+K^-$ invariant mass distribution of the decay $\Lambda_c
\to  pK^+K^-$ in $s$-wave with different values of $C=3,2,-2,-3$  and an arbitrary normalization factor $V_P$. The curves labeled as `both', `$I=0$', and `$I=1$' correspond to the contributions from the term of $t^{\rm s-wave}_{\Lambda_c\to pK^+K^-}$, $t^{I=0}$, and $t^{I=1}$, respectively.}
\label{Fig:dw_pKK_C}
\end{figure}

\begin{figure}[tbhp]
\begin{center}
\includegraphics[scale=0.6]{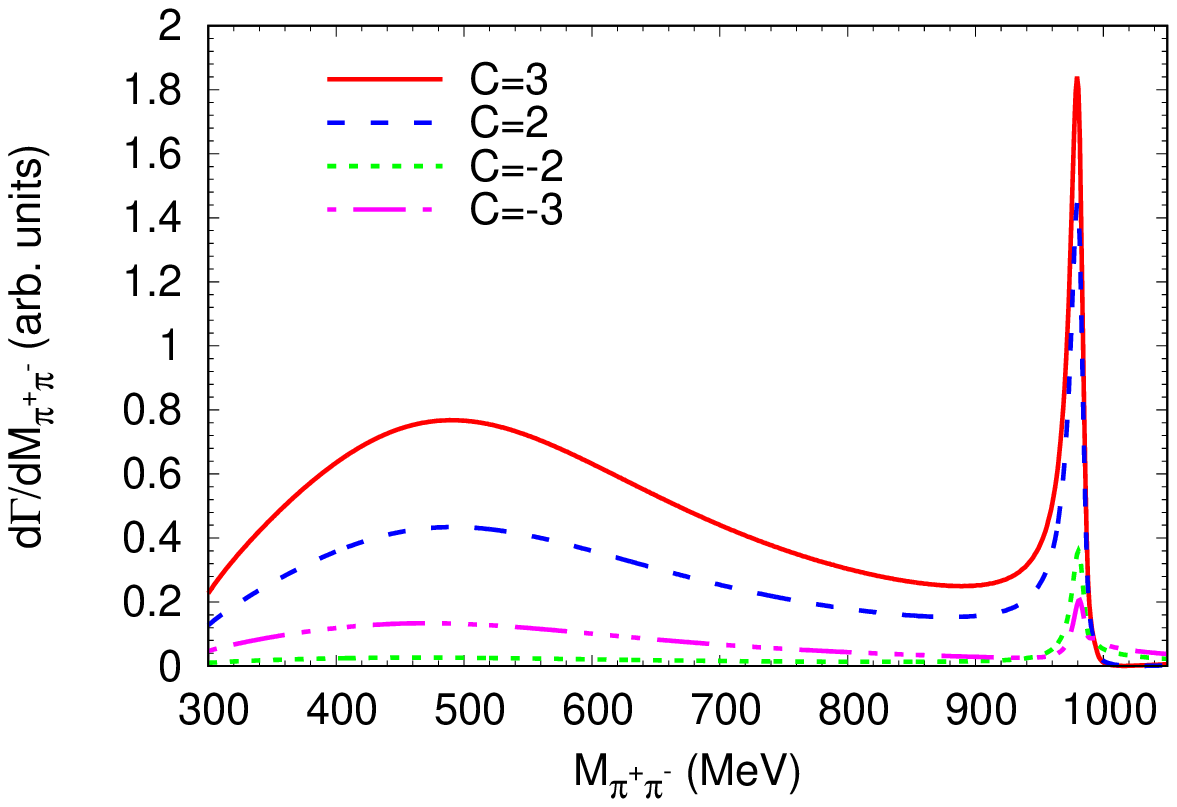}
\end{center}
\caption{The $\pi^+\pi^-$ invariant mass distribution of the decay $\Lambda_c
\to  p\pi^+\pi^-$ in $s$-wave with different values of $C=3,2,-2,-3$  and an arbitrary normalization factor $V_P$. }
\label{Fig:dw_ppipi_C}
\end{figure}

In addition to the factor $V_P$, we  have also the free parameter $C$, the relative weight of the external emission mechanism with respect to  the internal emission mechanisms. The value of $C$ should be around 3 because we take the number of the colors $N_c=3$, and the relative sign of $C$ is not fixed. We present the $K^+K^-$ and $\pi^+\pi^-$ invariant mass distributions with different values of $C=3,2,-2,-3$ in Figs.~\ref{Fig:dw_pKK_C} and \ref{Fig:dw_ppipi_C}, respectively.  In the $K^+K^-$ invariant mass distribution, one can see that the contributions from the isospin $I=1$ are much smaller than the ones of the isospin $I=0$ for the positive values of $C$, while both contributions from the isospin $I=0$ and $I=1$ are comparable for the negative values of $C$. This is because the coefficients of the terms $t^{I=0}$ and $t^{I=1}$ have the opposite sign before the $C$, and the contributions from the $\pi^0\eta$ and $(K\bar{K})_{I=1}$ [see Eq.~(\ref{eq:amp_f980_KK2})] have the negative interference for positive values of $C$. In both cases, one can find an enhancement structure close to the threshold, which is stronger for the positive values of $C$ and weaker for the negative values of $C$.  For the $\pi^+\pi^-$ invariant mass distribution of Fig.~\ref{Fig:dw_ppipi_C}, we can see a clear bump structure  around 500~MeV, and a sharp peak  around 980~MeV, which correspond to the $f_0(500)$ and $f_0(980)$, respectively. Both the signals are clearer for the positive value of $C$, and  weaker for the negative value of $C$. 

\begin{figure}[tbhp]
\begin{center}
\includegraphics[scale=0.6]{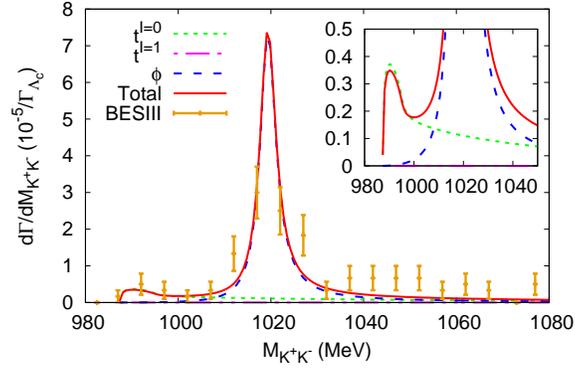}
\end{center}
\caption{The $K^+K$ invariant mass distribution of the $\Lambda_c
\to pK^+K^-$ decay  compared with the experimental
data from Ref.~\cite{Ablikim:2016tze}. The green dotted curve stands for the
contribution from the meson-meson interaction in $s$-wave, the blue
dashed curve corresponds to the results for the intermediate vector
$\phi$, and the red solid line shows the total contributions. }
\label{Fig:dw_pKK}
\end{figure}

\begin{figure}[tbhp]
\begin{center}
\includegraphics[scale=0.6]{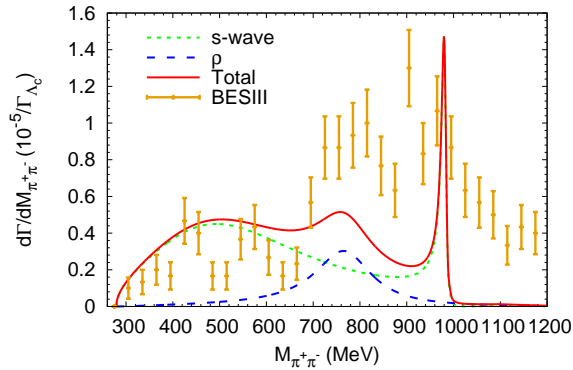}
\end{center}
\caption{The $\pi^+\pi^-$ invariant mass distributions of the
$\Lambda_c \to p \pi^+ \pi^-$ decay compared with the experimental
data from Ref.~\cite{Ablikim:2016tze}. The green dotted curve stands
for the contribution from the meson-meson interaction in $s$-wave,
the blue dashed curve corresponds to the results for the
intermediate vector $\rho^0$, and the red solid line shows the total
contributions.} \label{Fig:dw_ppipi}
\end{figure}

It should be stressed that although the BESIII Collaboration has reported the
$K^+K^-$ and $\pi^+\pi^-$ invariant mass distributions, we can not fit our model to to BESIII data which contain the background in the sideband region~\cite{Ablikim:2016tze}. In addition, we must bear in mind that the chiral unitary approach only makes reliable predictions up to 1100-1200~MeV. With the value of $(V'_P)^2/\Gamma_{\Lambda_c}=4.5\times
10^3$~MeV obtained above, we present the $K^+K^-$ and $\pi^+\pi^-$ invariant mass distributions by summing the contributions from the decays in $s$-wave and the intermediate vector mesons incoherently, as shown in Figs.~\ref{Fig:dw_pKK} and \ref{Fig:dw_ppipi}, respectively. For comparison, the BESIII data~\cite{Ablikim:2016tze} have been adjusted to the strength of our theoretical calculations.  We take the parameter $C=2$ and $(V_P)^2/\Gamma_{\Lambda_c}=0.2$~MeV$^{-1}$, in order to give rise to the sizeable signals of the $f_0(500)$ and $f_0(980)$~\footnote{It should be pointed out that the dimensions  of $V_P$ and $V'_P$ are `1' and `MeV' in our formalism, respectively,  thus one can not obtain the relative weight of the three-body decay and two-body decay of the $\Lambda_c$ by comparing the $V_P$ with $V'_P$. }. Both the parameters can be obtained by fitting to the experimental data, when more precise measurement of the processes is available in future. For the $K^+K^-$ invariant mass distribution, our model produces an enhancement structure close to the threshold mainly due to the resonance $f_0(980)$, and a clear peak of the $\phi$, which are in good agreement with the BESIII data.
It is worth mentioning that, in the $K^+K^-$ invariant mass distribution of the decay
$\chi_{cJ}\to p\bar{p}K^+K^-$ measured by the BESIII
Collaboration~\cite{Ablikim:2011uf}, one can find an enhancement
structure close to the threshold, which can be associated to the
resonance $f_0(980)$ and $a_0(980)$. A similar structure can also be found in the
decay $D^+_s \to K^+K^-\pi^+$ measured by the BABAR Collaboration~\cite{delAmoSanchez:2010yp}.

For the $\pi^+\pi^-$ invariant mass
distribution of the decay $\Lambda_c\to p \pi^+\pi^-$ as shown in
Fig.~\ref{Fig:dw_ppipi}, one can see a clear peak around
770~MeV, corresponding to the vector meson $\rho^0$, and a broad peak
around 500~MeV, which can be associated to the scalar meson
$f_0(500)$, dynamically generated from the meson-meson interactions
in $s$-wave. In addition, there is a narrow sharp peak around 980~MeV for the scalar state $f_0(980)$. We can see that the broad peak for $f_0(500)$, the peak
for $\rho^0$, and a narrow sharp one for $f_0(980)$~\footnote{This peak would be a bit broader with the peak strength reduced if it is folded with the experimental resolution.} of our results are
compatible with the BESIII measurement~\cite{Ablikim:2016tze}.

From Figs.~\ref{Fig:dw_pKK} and \ref{Fig:dw_ppipi}, one can find that the results with $C=2$ are in reasonable agreement with the BESIII measurements~\cite{Ablikim:2016tze}, which implies that the $W$ external emission mechanism is more important than the $W$ internal emission mechanism. According to the topological classification of the weak decays in Refs.~\cite{Chau:1982da,Chau:1987tk}, the strength of $W$ external emission is larger than the one of $W$ internal emission. It should be stressed that our results strongly depend on the sign of $C$, and the present measurements of the BESIII Collaboration favor $C=2$ and a much smaller contribution from the $a_0(980)$. Indeed, if there is a sizeable contribution from the $a_0(980)$ in $\Lambda_c \to p K^+K^-$, it implies that we can observe the process $\Lambda_c\to p \pi^0 \eta$, and the signal of the $a_0(980)$ in the $\pi^0\eta$ mass distribution experimentally, however there are no any report about this process~\cite{PDG2018}.

\section{Conclusions}
\label{sec:conc}

In this work, we have studied the decays $\Lambda_c\to p K^+K^-$ and
$\Lambda_c\to p \pi^+\pi^-$, by taking into account contributions of
the intermediate vector mesons, and the $s$-wave meson-meson
interactions within the chiral unitary approach, where the
$f_0(500)$, $f_0(980)$, and $a_0(980)$ resonances are dynamically generated.

The $K^+K^-$ and $\pi^+\pi^-$ invariant mass distributions for these
two decays are calculated. In the $K^+K^-$ invariant mass distribution, one
can find a narrow peak for the $\phi$, and an enhancement structure
close to the $K^+K^-$ threshold, which should be the reflection of the
$f_0(980)$ and $a_0(980)$ resonances.  For the $\Lambda_c\to p \pi^+\pi^-$ mass distribution, in
addition to the broad peak of the $\rho^0$, one can find a bump
structure around 500~MeV for the $f_0(500)$, and a narrow sharp peak
around 980~MeV for the $f_0(980)$, in agreement with
the BESIII measurement.

According to our calculations, the present measurements of the BESIII Collaboration favor a much smaller contribution from the $a_0(980)$. As we discussed, if there is a sizeable contribution from the $a_0(980)$ in $\Lambda_c \to p K^+K^-$, it implies that we can observe the process $\Lambda_c\to p \pi^0 \eta$, and the signal of the $a_0(980)$ in the $\pi^0\eta$ mass distribution experimentally, however there are no any report about this process~\cite{PDG2018}.

We encourage our experimental colleagues to measure these two decays, which can be used to test the molecular
nature of the scalar resonances $f_0(500)$, $f_0(980)$, and $a_0(980)$.

\section*{Acknowledgments}
We warmly thank Hai-Ping Peng, Ya-Teng Zhang, and Yue Pan for sending us the BESIII experimental data files, and thank Eulogio Oset for carefully reading.
This work is partly supported by the
National Natural Science Foundation of China under Grants Nos.
11505158, 11947089, 11735003 and 11961141012. It is also supported by
the Key Research Projects of Henan Higher Education Institutions
under No. 20A140027, the Academic Improvement Project of Zhengzhou
University, and the Youth Innovation Promotion Association CAS
(2016367).

  \end{document}